\documentclass[sigconf]{acmart}

\usepackage{multirow}
\usepackage{colortbl}
\usepackage{bm}
\usepackage{enumitem}
\usepackage{mathtools}

\usepackage{float}
\usepackage{listings}

\renewcommand\footnotetextcopyrightpermission[1]{} % removes footnote with conference information in first column
\acmBooktitle{}

\newcommand{\MethodName}{\textbf{\textit{RubricRAG}}}
\usepackage{fancyvrb}
\usepackage{todonotes}

\begin{document}

\title{\textit{RubricRAG}: Towards Interpretable and Reliable LLM Evaluation via Domain Knowledge Retrieval for Rubric Generation}

\author{Kaustubh D. Dhole}
\affiliation{%
  \institution{Department of Computer Science}
  \institution{Emory University}
  \city{Atlanta}
\state{GA}
 \country{USA}
  }
\email{kdhole@emory.edu}

\author{Eugene Agichtein}
\affiliation{%
  \institution{Department of Computer Science}
  \institution{Emory University}
  \city{Atlanta}
\state{GA}
 \country{USA}
  }
\email{eugene.agichtein@emory.edu}

\renewcommand{\shortauthors}{Dhole and Agichtein}

\begin{abstract}
Large language models (LLMs) are increasingly evaluated and sometimes trained using automated graders such as LLM-as-judges that output scalar scores or preferences. While convenient, these approaches are often opaque: a single score rarely explains why an answer is good or bad, which requirements were missed, or how a system should be improved. This lack of interpretability limits their usefulness for model development, dataset curation, and high-stakes deployment. Query-specific rubric-based evaluation offers a more transparent alternative by decomposing quality into explicit, checkable criteria. However, manually designing high-quality, query-specific rubrics is labor-intensive and cognitively demanding and not feasible for deployment. While previous approaches have focused on generating intermediate rubrics for automated downstream evaluation, it is unclear if these rubrics are both interpretable and effective for human users. In this work, we investigate whether LLMs can generate useful, instance-specific rubrics as compared to human-authored rubrics, while also improving effectiveness for identifying good responses. Through our systematic study on two rubric benchmarks, and on multiple few-shot and post-training strategies, we find that off-the-shelf LLMs produce rubrics that are poorly aligned with human-authored ones. We introduce a simple strategy,~\MethodName, which retrieves domain knowledge via rubrics at inference time from related queries. We demonstrate that \MethodName{} can generate more interpretable rubrics both for similarity to human-authored rubrics, and for improved downstream evaluation effectiveness. Our results highlight both the challenges and a promising approach of scalable, interpretable evaluation through automated rubric generation.
\end{abstract}


\maketitle

\section{Introduction and Background}

\begin{figure}
    \centering
    \includegraphics[width=1\linewidth]{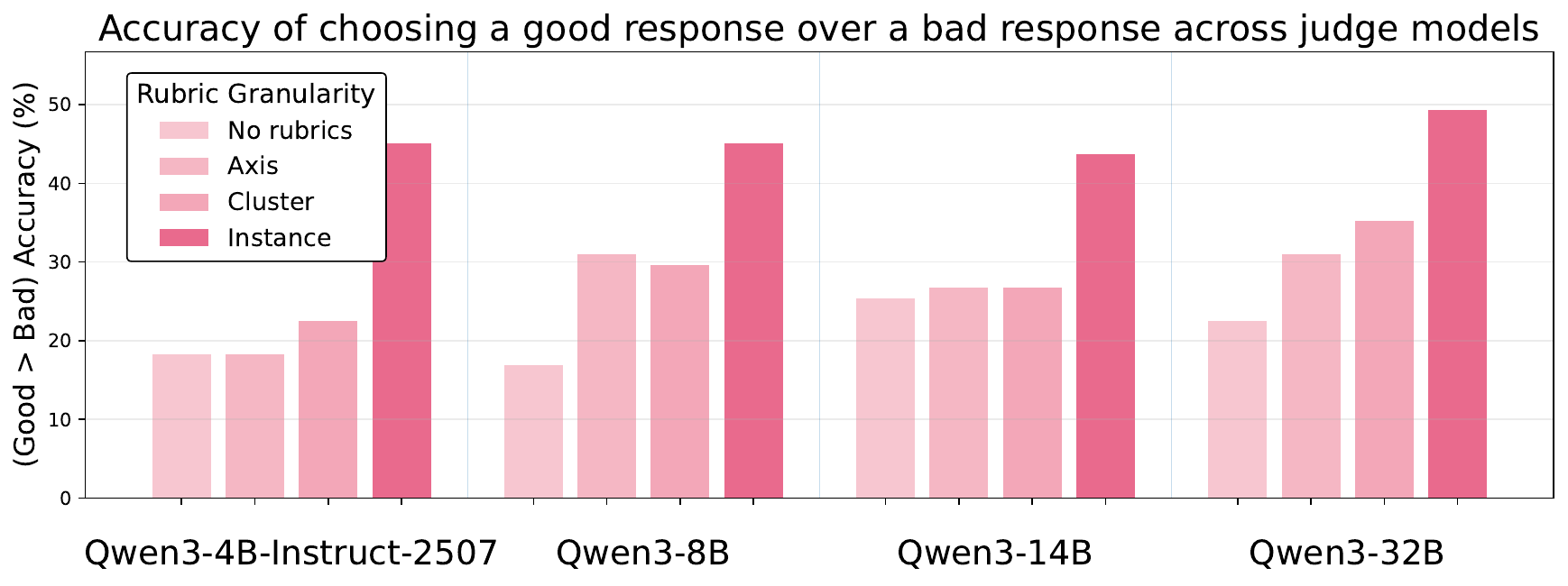}
    \caption{Fine-grained rubrics consistently show higher accuracy in preferring good over bad responses. Moreover, interpretable evaluations using both cluster- and instance-level rubrics outperform evaluations without rubrics.}
    \label{fig:cluster_vs_instance}
\end{figure}

Large language models (LLMs) are increasingly evaluated, and in many settings, even trained, using automated graders, like LLM-as-judges that generally output a preference or a scalar score~\cite{dubois2024alpacafarmsimulationframeworkmethods,srivastava2023beyond,dhole2025conqret,es-etal-2024-ragas, saad-falcon-etal-2024-ares}. While such LLM-as-judge pipelines are convenient, they are often opaque: a single number rarely explains \emph{why} an answer is good or bad, what specific requirements were missed, or how to improve a system over successive iterations~\citep{ye-etal-2025-tooleyes, kim-etal-2025-biggen}. This lack of interpretability complicates model development, dataset curation, and deployment in high-stakes domains where specific actionable feedback matters, like addressing sensitive health queries.

Rubric-based evaluation~\cite{jonsson2007use,brookhart2013create,min-etal-2023-factscore,kim2023prometheus,dhole2025conqret}, on the other hand, decomposes an otherwise fuzzy notion of ``quality'' into explicit, checkable criteria (e.g., factual correctness, citation support, safety constraints, completeness, tone) enabling fine-grained diagnostics and feedback~\cite{farzi2024pencils,dhole2025adversem,ye2024flask,feng-etal-2025-mad}. Although intended across the dataset, such criteria may be too vague to capture the specific requirements of individual queries, resulting in less effective evaluation. 

Query-specific rubrics, rather than generalizing the evaluation of all query types through common criteria, allow for gauging the particular requirements of individual queries. Such specificity can be useful for interpretability as well as downstream LLM evaluation. For instance, on a subset of queries from OpenAI HealthBench~\cite{arora2025healthbench}, we find that multiple LLM judges from the Qwen family~\cite{yang2025qwen3} are more effective at choosing good over bad responses when supported by fine-grained, query-specific rubrics than when supported by generalized, coarse-level rubrics, or even no rubrics (Figure~\ref{fig:cluster_vs_instance}).

Fine-grained rubrics have been focused across diverse domains~\cite{fan-etal-2024-sedareval,dhole2025conqret}: Recently OpenAI HealthBench introduced physician-written, query-specific rubrics for medical dialogues~\citep{arora2025healthbench}; ResearchRubrics~\cite{sharma2025researchrubrics} designed structured, instance-level criteria for deep research tasks. Apart from supporting downstream judges, fine-grained rubrics have been effective as structured reward signals for reinforcement learning in settings without strict verification, outperforming scalar rewards~\cite{gunjal2025rubricsrewardsreinforcementlearning,shao2025dr,liu2025openrubrics,li2026rubrichub}.

While there are a plethora of benefits of fine-grained rubrics, it is hard to obtain them at scale. Even for a domain expert, designing high-quality rubrics for each query can be tedious and cognitively demanding. It requires analyzing different dimensions, appropriate granularity of evaluation, and often, different trade-offs between accuracy and safety~\cite{jonsson2007use,brookhart2013create}.

In this work, we investigate whether LLMs themselves can help overcome this bottleneck by~\emph{automatically generating} useful, fine-grained human-like rubrics. LLMs have broad world knowledge and exposure to many genres of instruction and assessment~\cite{kazemi2025big,phan2025humanity}, suggesting they may be capable of proposing evaluation dimensions that are both comprehensive and actionable. 

Specifically, we ask the following questions: \textit{\textbf{RQ1}: Can LLMs generate fine-grained query-specific rubrics that are similar to human-authored rubrics?} \textit{\textbf{RQ2}: Can such LLM-generated rubrics be useful for downstream evaluation to choose good over bad responses?}

In that regard, our contributions are as follows:
(i) We first introduce three rubric-generation evaluation metrics -- \textbf{\textit{Rubric-BLEU}}, \textbf{\textit{Rubric-ROUGE}}, and a \textbf{\textit{Rubric-LLM-judge}} to quantify the alignment with human rubrics under both lexical and semantic criteria.
(ii) We then evaluate multiple approaches of employing LLMs for rubric generation. We find that when prompted in a zero-shot fashion, LLMs are poor rubric generators.
(iii) To generate human-like rubrics, we introduce~\textbf\textit{{\MethodName}} and show how retrieving rubrics from similar queries can be extremely effective. We also show how two popular post-training approaches, namely, supervised fine-tuning (SFT) and a group relative policy optimization (GRPO) based reinforcement learning (RL) approach trained using multi-objective rewards, can also improve rubric generation abilities. 
(iv) We demonstrate that retrieval-augmented rubric generation improves downstream evaluation quality, yielding stronger alignment with human-rubric-based judgments and better discriminative power between good and bad responses.

This paper is organized as follows. In Section~\ref{sec:rel_work}, we first discuss related work. In Section~\ref{sec:methods} we present the task and the methods employed. In Section~\ref{sec:results}, we present the evaluation of generated rubrics, and conclude in Section~\ref{sec:conclusion}.
\section{Related Work}\label{sec:rel_work}
\paragraph{Fine-Grained LLM Judges.}
Traditional LLM-as-judge pipelines often rely on a single preference or scalar score~\cite{kaufmann2024survey}, which can obscure specific strengths and weaknesses, especially for long-form or high-stakes responses. A growing body of work argues that decomposing evaluation into explicit dimensions improves downstream evaluation~\cite{dhole2025conqret, 10.1145/3664190.3672511,dhole-etal-2025-generative}. For example, FLASK evaluates alignment through fine-grained skill sets, showing improved agreement with human judgments compared to coarse scores~\cite{ye2024flask}. Similarly, multi-dimensional evaluation frameworks such as M-MAD demonstrate that scoring across separate criteria yields more robust and accurate judgments than single aggregated scores~\cite{feng-etal-2025-mad}. In long-form retrieval-augmented settings, ConQRET show that task-specific fine-grained rubrics are effective for answer quality evaluation~\cite{dhole2025conqret}, while AdverSEM~\cite{dhole2025adversem} uses structured perturbations to evaluate factual robustness across multiple dimensions. Across these settings, fine-grained criteria consistently provide more interpretable and reliable assessments than coarse scoring.
\paragraph{Query-Specific Rubrics for Evaluation.} 
A complementary direction structures evaluation as a checklist of verifiable items. RocketEval reframes judging as answering a set of checklist questions about an output, enabling small evaluator models to achieve high correlation with human preferences \citep{weirocketeval}.
\paragraph{Rubrics as training signals beyond verifiable tasks.}
Beyond an evaluation artifact, rubrics can also shape learning by providing multi-faceted feedback~\cite{mu2024rule,huang2025reinforcement,10.1145/3664646.3664778,zhang2026rubricbench}. Rubrics as Rewards~\cite{gunjal2025rubricsrewardsreinforcementlearning} proposes rubrics as reward signals for reinforcement learning in domains where strict verification is difficult, demonstrating gains over scalar reward formulations and has been adopted in various works~\citep{shao2025dr,liu2025openrubrics,li2026rubrichub,goel2025training,huang2026rubicap,argo2026interpretingblackboxrewardmodels,viswanathan2025checklists}.
\section{Methods and Experiments}\label{sec:methods}
We now describe the task and our rubric generation approaches.
\subsection{Task: Rubric generation}\label{sec:task}
Given a user query $q$, we are interested to generate a set of fine-grained rubrics $R = {r_1, r_2 \ldots }$ that can be used to grade the assistant's next response. Each set of rubrics is a list of criteria, where each criterion is paired with an integer point value. Positive points reward desirable behavior (e.g., clinically correct advice, safe triage, clear communication), while negative points penalize failure modes (e.g., unsafe instructions, missed red flags, hallucinated medical claims). Criteria may be either positive or negative and are intended to cover both \emph{what to do} and \emph{what to avoid}.

\subsection{Rubric Generation Approaches}\label{sec:prompting}
We would like to see how LLMs ($M_\theta$) perform both in a raw zero-shot fashion, generally employed in agentic style workflows, as well as measure how we can provide additional useful context from other queries to be able to generate effective rubrics $\hat{R}$. 
\begin{equation*}
\hat{R}=M_\theta(q),
\end{equation*}

\paragraph{\textbf{Zero-shot and Few-Shot rubric generation.}}
In the zero-shot setting, the generator simulates the role of an annotator, where we provide instructions to produce a list of rubrics in a strict JSON format, including both positive and negative criteria with integer point weights. We additionally prepend $k$ random exemplar pairs from the training set $(q^{(j)},R^{(j)})$ to gauge few-shot performance.

\paragraph{\textbf{Retrieving from similar queries (RubricRAG)}}
Here, we use the user query to retrieve $k$ similar queries from the training set,
using a dense retriever $\phi$, and incorporate their corresponding rubrics as few-shot exemplars
$ \phi(q) = \{(q^{(j)},R^{(j)})\}_{j=1}^{k}$ pairs in context. The main motivation of retrieving similar queries is to familiarize the model beyond the domain and task with understanding query-specific nuances. 

\begin{figure}[!htbp]
\centering
\begin{minipage}{\columnwidth}
\scriptsize
\begin{Verbatim}[frame=single]
You are a physician-annotator creating evaluation rubrics for AI health 
assistants.
Given a health-related conversation, you must write a set of rubric criterion 
that can be used to grade a final assistant response. Each rubric criterion 
describes a specific requirement or failure mode and assigns it an integer 
point value. Positive points reward desirable behavior; negative points 
penalize unsafe, incorrect, or otherwise harmful behavior. 
Base your rubrics ONLY on the conversation content and what matters clinically.
{Query}
Task: Generate a comprehensive set of positive and negative rubric criterion 
that would be used to grade an AI assistant's next response to 
this conversation.
Given the above query, you must output evaluation rubrics in the following 
strict JSON format:
{
  "rubrics": [
    {
      "criterion": "<description of what the model should or should not do>",
      "points": <integer, can be positive or negative>
    },
    ...
  ]
}
Guidelines:
- Include BOTH positive criterion (things a good response SHOULD do, 
with positive points) and negative criterion (things a response MUST AVOID 
or where failure should be penalized, with negative points).
- Positive criterion should capture clinically important behaviors:
  safety, correctness, communication quality, context-aware triage, 
  uncertainty handling, etc.
- Negative criterion should punish dangerous, misleading, or incomplete 
behaviors:  unsafe advice, overconfidence, ignoring red flags, failing to 
escalate care, etc.
- Points should roughly encode importance:
  higher positive points for critical requirements; more negative points for 
  severe failures.
- Do NOT include any fields other than "criterion" and "points".
- Do NOT include comments or trailing commas.
- Make sure the JSON is syntactically valid. Your JSON format should 
be strictly followed.
\end{Verbatim}
\end{minipage}
\caption{System and user prompt used to generate health-assistant evaluation rubrics.}
\label{fig:rubric-prompt}
\end{figure}

\begin{figure}[!htbp]
\centering
\begin{minipage}{\columnwidth}
\scriptsize
\begin{Verbatim}[frame=single]
You are an expert evaluator of rubric criterion similarity for health AI
systems. Given ONE reference criterion and ONE generated criterion, output 
a single integer score in [0,9] representing semantic similarity 
(9 = same meaning, 0 = unrelated).
Output ONLY the number. No JSON. No explanation. No extra text.
REFERENCE: {ref_text} GENERATED: {gen_text} Similarity score (0..9):
\end{Verbatim}
\end{minipage}
\caption{LLM judge criterion similarity prompt.}
\label{fig:llmjudge-prompt}
\end{figure}

\paragraph{\textbf{Supervised fine-tuning (SFT)}} Here, we fine-tune the generator to directly predict the human-authored rubrics conditioned on the user query, using teacher forcing with a causal language modeling objective over the concatenated prompt-and-target sequence. We use (Q)LoRA adapters to reduce trainable parameters. 

\paragraph{\textbf{GRPO with multi-objective rewards.}}
We also introduce an RL-based approach where we optimize the generator with Group Relative Policy Optimization (GRPO)~\cite{shao2024deepseekmath} using sparse weighted rewards. Here, the generator acts as a policy and generates reasoning steps before generating the final rubrics. We reward the policy's rollout as a weighted sum of four reward functions that apply over the generated rubrics -- (i) binary format correctness ($r_{\text{f}}$), (ii) similarity with human-authored rubrics ($r_{\text{s}}$), (iii) diversity among generated rubrics ($r_{\text{d}}$), (iv) and normalized deviation of mean and variance of generated rubrics from the reference rubrics ($r_{\text{l}}$):
\begin{equation*}
\mathcal{R} = w_{\text{f}}\,r_{\text{f}} + w_{\text{s}}\,r_{\text{s}} + w_{\text{d}}\,r_{\text{d}} + w_{\text{l}}\,r_{\text{l}}
\end{equation*}

\subsection{Evaluation metrics}\label{sec:metrics}
We measure the quality of the rubrics using our rubric similarity metrics and two downstream evaluations over a fixed LLM-judge.

\subsubsection{\textbf{Rubric Similarity Metrics}}\label{subsec:rubricsim_metrics}
Standard generation metrics like BLEU~\cite{papineni2002bleu} and ROUGE~\cite{lin-2004-rouge} are typically computed at the corpus or full-text level, but in our setting, we instead define a macro-averaged, per-criterion ``best overlap'' where the generated rubric is treated as a set of criteria rather than a single string. We compute them in both directions—generated-to-reference (precision) and reference-to-generated (recall). Our formulation is~\textbf{permutation invariant} and is able to evaluate each of the criteria with respect to reference criteria without preferring any ordering between them.  In addition to n-gram overlap, we also use a lightweight LLM judge~\cite{dhole2024llm, dhole2025conqret} to capture semantic similarity. 

% TODO (Evaluation metrics): define both metrics (P/R/F1) with equation; describe max/all; harmonic mean
Let $R=\{c_i\}_{i=1}^{m}$ be gold criteria and $\hat{R}=\{\hat{c}_j\}_{j=1}^{n}$ be generated criteria. For a similarity function $s(\cdot,\cdot)\in[0,1]$ (e.g., ROUGE score), we define the corresponding rubric similarity metric, viz., \textbf{\textit{Rubric-BLEU}}, \textbf{\textit{Rubric-ROUGE}}, and \textbf{\textit{Rubric-LLM-Judge}} as follows:
\begin{align*}
P &= \frac{1}{n}\sum_{j=1}^{n}\max_{i\in[m]} s(\hat{c}_j,c_i), & R &= \frac{1}{m}\sum_{i=1}^{m}\max_{j\in[n]} s(c_i,\hat{c}_j), & F_1 &= \frac{2PR}{P+R}\\
\end{align*}
%We report Rubric-BLEU, Rubric-ROUGE, and analogously for the LLM-judge score as F1 scores.

\subsubsection{\textbf{Hallucinations, Misses and Redundant Rubrics}}\label{subsec:hall_andmissed}
We additionally track the propensity for hallucinations among the generated rubrics, the percentage of rubrics missed, and the redundancy among generated rubrics through the following query-wise metrics. Let $R=\{c_i\}_{i=1}^{m}$ denote the reference rubrics and $\hat{R}=\{\hat{c}_j\}_{j=1}^{n}$ denote the generated rubrics. Let $s(\cdot,\cdot)\in[0,1]$ be a similarity function, and let $\mathbf{1}[\cdot]$ denote the indicator function.
\\ \\
\noindent We define 
\textbf{Missed@}$\tau$  to measure the fraction of reference rubrics that are not sufficiently covered by any generated rubric, 
\begin{align*}
\textbf{Missed@}\tau 
&= \frac{1}{m}\sum_{i=1}^{m} \mathbf{1}\!\left[\max_{j\in[n]} s(c_i,\hat{c}_j) < \tau \right]
\end{align*}
\textbf{Hallucinations@}$\tau$ to measure the fraction of generated rubrics that do not sufficiently match any reference rubric,\begin{align*}
\textbf{Hallucinations@}\tau 
&= \frac{1}{n}\sum_{j=1}^{n} \mathbf{1}\!\left[\max_{i\in[m]} s(\hat{c}_j,c_i) < \tau \right]
\end{align*}
and \textbf{Redundancy@}$\tau$ measures the fraction of generated rubric pairs that are overly similar to each other.

\begin{align*}
\textbf{Redundancy@}\tau 
&= \frac{2}{n(n-1)}\sum_{1 \le j < k \le n} \mathbf{1}\!\left[s(\hat{c}_j,\hat{c}_k) > \tau \right]
\end{align*}
\noindent

In addition to the above intrinsic metrics, we also perform downstream evaluations using LLM judges:

\subsubsection{\textbf{Downstream Rubric Utility}}\label{subsec:downstream}

We evaluate the downstream effectiveness of the generated rubrics in two settings: 

i) \textbf{Query-wise Correlation of LLM-Judge Scores Obtained From Model-Generated and Human-Authored Rubrics} \\
Here, we use an LLM judge in the style of HealthBench~\cite{arora2025healthbench}. For each query, the judge evaluates a human-authored response against each rubric criterion individually, producing a binary yes/no decision. Points for all satisfied criteria are summed to obtain a query-level score and normalised by dividing with the sum of all positive criterion. We do the same using the human-authored (gold) rubrics as well and measure the correlation between the two sets of scores, as well as the dataset level average scores.\footnote{We validate this with human-authored rubrics, and our LLM judge: We obtain a score of .37 which is in the range of HealthBench's analysis of closed-sourced models.} 

ii) \textbf{Ability to Prefer Good Response Over Bad Response} In addition to such pointwise correlations, we also evaluate the discriminative potential of the rubrics to prefer good response against bad ones. For good responses, we use the human-authored completions provided by HealthBench, while for bad responses, we force an LLM to generate a response by adhering to rubrics associated with other random queries. We describe the details in the following section.
\subsection{Evaluation Across Several Rubric Granularities}\label{appendix:insta_vs_clus}
Before employing models for generating rubrics, we wanted to know whether human-authored rubrics of different granularities themselves benefit LLM-Judges to discriminate good from bad responses better than no rubrics at all.

Specifically, we gauge whether fine-grained rubrics are more effective than coarse-level global rubrics at preferring good over bad responses, by evaluating various models of different sizes, on four settings: 1) \textbf{\textit{No rubrics}}, 2) \textbf{\textit{Axis-level rubrics}}, which consist of 5 static rubrics (accuracy, communication quality, completeness, context awareness, and instruction following ability) 3) \textbf{\textit{Cluster-level}} rubrics (consisting of 37 rubrics which are shared across many queries) which are more fine-grained than axis-level rubrics but are shared across queries 4) \textbf{\textit{Query-specific rubrics}} (where each query-completion is evaluated with rubrics specific to the query's context. Axes and clusters have been computed by the authors of HealthBench~\cite{arora2025healthbench}.
\paragraph{\textbf{Model Performance across Granularities}} 
We then score how well different models, acting as LLM Judges, prefer the good response across each of the four granularities of rubrics.
Our rubric evaluation approach is similar to the one performed by~\citet{arora2025healthbench}. For each granularity, the LLM Judge decides whether every rubric (criterion) is satisfied by the good and bad responses separately. Each rubric is prompted one at a time. The sum of the points of the satisfied criterion is treated as the score of the response. For the no rubric setting, we prompt the LLM Judge to output a single score. This score is further normalized by dividing by the points of the positive rubrics. 

\paragraph{{\textbf{Creating Good versus Bad Responses}}} 
To create an evaluation set of good versus bad completions, we gather the physician-written completions and treat them as good responses. For gathering bad completions, we prompt a~\texttt{Qwen3-30B-A3B-Instruct-2507} model with a HealthBench query alongwith rubrics from other random queries, and instruct the model to generate a response conditioned on those rubrics. We release the model completions on HuggingFace~\cite{lhoest2021datasets} at:~\href{https://hf.co/datasets/kdhole/healthbench-rubric-responses}{\texttt{\textbf{kdhole/healthbench-rubric-responses}}}.

\subsection{Experimental Setup}\label{sec:models}
We use~\texttt{Qwen3-14B}\footnote{We investigated smaller LMs like~\texttt{Qwen3-0.6B, 1.7B, 4B-Instruct, and 8B} and found frequent malformed JSONs, and would require significant output cleaning logic.} as the rubric generator. The prompts used for generation and downstream rubric evaluation are shown in Figures~\ref{fig:rubric-prompt} and~\ref{fig:llmjudge-prompt}, respectively. For judging rubric entailment to compute \textbf{\textit{Rubric-LLM-JUDGE}}, and downstream evaluation, we use~\texttt{Qwen3-4B-Instruct-2507}\footnote{Note that if the rubric entailment task is framed in different ways like generating all rubrics at a time, a larger model may be needed.}. In our experiments, we use $k=20$ exemplars for HealthBench and $k=5$ for ResearchRubrics.
We use greedy decoding and a maximum token length of 1024; unless stated otherwise, we enable the model's \texttt{thinking} mode in the chat template during generation. For SFT, we train with LoRA adapters (rank $r=16$, $\alpha=32$, dropout $0.05$) using learning rate $5e{-5}$ and disable \texttt{thinking} mode. For GRPO, we set reward weights $(w_{\text{fmt}} = 1,w_{\text{sim}} = 5,w_{\text{div}} = 2,w_{\text{len}} = 1)$, and implement training with HuggingFace Transformers~\cite{wolf2020transformers} and the Transformers Reinforcement Learning~\cite{vonwerra2020trl} libraries. For RubricRAG, we investigate two settings, with and without intermediate thinking tokens, RubricRAG (think) and RubricRAG (nothink). For retrieving similar queries, we resort to~\texttt{Qwen3-Embedding-4B}~\cite{qwen3embedding} using the Sentence Transformers library~\cite{reimers-2019-sentence-bert}.

\subsection{Datasets and splits}\label{sec:data}
We report rubric generation performance on three evaluation sets. We use the OpenAI HealthBench dataset as it contains a large number of queries with rubrics written by human experts. Each example contains (i) a complex user query; and (ii) a reference rubric list (\texttt{rubrics}) authored by physicians. We use 300 random queries from their \texttt{oss\_eval} subset, and all queries from the \texttt{hard} subset for evaluation, and the remaining queries in \texttt{oss\_eval} are used for training. Additionally, we report rubric generation performance on the ResearchRubrics dataset as well, which contains 101 queries, with fine-grained rubrics. Since this dataset is small, we set aside 5 queries for few-shot examples, and the remaining as the evaluation set. To evaluate RubricRAG on this dataset, we allow searching from all other queries except for the test query.

\begin{table*}[t]
\centering
\resizebox{\textwidth}{!}{%
\begin{tabular}{l |r r r r r  |r r r r r|r r r r r}
\toprule
& \multicolumn{5}{c|}{\textbf{HEALTHBENCH (OSS EVAL-300)}} & \multicolumn{5}{c}{\textbf{HEALTHBENCH (HARD)}} &  \multicolumn{5}{c}{\textbf{RESEARCH RUBRICS}}\\
%\midrule
%\cmidrule(lr){2-6}\cmidrule(lr){7-12}
\textbf{MODE} 
& \textbf{BLEU} 
& \textbf{ROUGE1} 
& \textbf{ROUGE2} 
& \textbf{ROUGEL} 
& \textbf{LLM-JUDGE} 
& \textbf{BLEU} 
& \textbf{ROUGE1} 
& \textbf{ROUGE2} 
& \textbf{ROUGEL} 
& \textbf{LLM-JUDGE} & \textbf{BLEU} 
& \textbf{ROUGE1} 
& \textbf{ROUGE2} 
& \textbf{ROUGEL} 
& \textbf{LLM-JUDGE}\\
\midrule
Zero-Shot 
& .020  &.231 & .065 & .192 & .521 
& .015  & .216 & .057 & .177 & .474  
& .092&.224&.071&.180& .514\\

Few-Shot 
& .033  & .293 & .091 & .237 & .548 
& .029  & .277 & .083 & .220 & .505  & .133& .335& .129& .275& .501\\

GRPO & .035 & .311 & .101 & .248 & .551 &
.027 & .291 & .089 & .231 & .506  & – & – & – & – & –\\

SFT & .042 & .311 & .103 & .253 & .481
& \textbf{.042} & .299 & .098 & .241 & .436 & – & – & – & – & –\\
RubricRAG (think) & .037  & .304 & .097 & .245 & \textbf{.558}  & .030  & .283 & .085 & .225 & .514  & \textbf{.136}& .337& \textbf{.135}& .283 & .523\\

RubricRAG (nothink)  
& \textbf{.049}  & \textbf{.331} & \textbf{.115} & \textbf{.269} & .567 
& .039 & \textbf{.311} & \textbf{.103} & \textbf{.251 } & \textbf{.523}  & .089& \textbf{.339}& .133& .286& \textbf{.576}\\

%Diversified RubricRAG & .035 & \textbf{.304} & \textbf{.245} & \textbf{.559} & \textbf{.030} & \textbf{.284} & \textbf{.227} & \textbf{.515} \\
\bottomrule
\end{tabular}%
}
\caption{Rubric Generation Performance of Qwen3-14B on OpenAI HealthBench~\cite{arora2025healthbench}. 
All values are Rubric-$\ast$ F1 scores. 
SFT and GRPO were not evaluated on ResearchRubrics~\cite{sharma2025researchrubrics} due to the absence of a training set. }
\label{tab:model_comparison}
\end{table*}

\section{Results}\label{sec:results}
We now present the results of our experiments. 
\subsection{Downstream Effectiveness of Different Granularities of Human-Authored Rubrics}
We first gauge whether human-authored rubrics at various granularities themselves help in evaluation.

\noindent\paragraph{\textbf{Fine-Grained Rubrics are more Discriminative}}
We find that query-specific human-authored rubrics consistently show higher accuracy in preferring the human-written (good) response over the response generated with randomly conditioned rubrics (bad), as shown in Figure~\ref{fig:cluster_vs_instance}. Moreover, \textbf{interpretable evaluations using both cluster- and instance-level rubrics outperform evaluations without rubrics}. Besides, we also find that rubrics that are coarser are marked as satisfied for both good and bad completions by the LLM Judges, resulting in many ties.  

We now discuss the results for model-generated query-specific rubrics.
\subsection{Similarity to Human-Authored Rubrics.}
Table~\ref{tab:model_comparison} shows the similarity between generated rubrics and human-authored rubrics across the three evaluation sets. Overall, we observe that using off-the-shelf LLMs result in poor rubric generators in the zero-shot setting. Across all the three benchmarks, zero-shot performance is consistently low on rubric-BLEU and rubric-ROUGE, and only moderate on LLM-judge evaluation. This suggests that while models may capture some high-level intent, they fail to reproduce the fine-grained structure and clinically grounded criteria present in human-authored rubrics.

Providing random few-shot exemplars improves performance across all metrics. Even randomly sampled examples lead to noticeable gains in both lexical overlap and semantic similarity, indicating that models benefit from seeing the expected rubric format, level of granularity, and balance of positive and negative criteria. However, the improvements remain modest, suggesting that simple few-shot prompting is insufficient to reliably produce human-like rubrics.

Retrieval augmented rubric generation further improves alignment. When exemplars are selected using retrieval over similar queries, performance increases across nearly all metrics. In particular, the RubricRAG approach achieves the highest rubric-similarity scores, indicating that semantically similar examples help the model produce more human-style rubrics. Notably, this simple retrieval strategy performs better than expensive post-training approaches like SFT which require immense training data.

Post-training methods generate better rubrics than zero-shot and few-shot approaches, where supervised fine tuned approach outperforming all. Supervised fine-tuning (SFT) produces strong lexical similarity scores, achieving high rubric-BLEU and rubric-ROUGE on both evaluation sets. However, it underperforms on the semantic LLM-judge metric, suggesting that improved surface overlap does not necessarily translate to better semantic alignment. The GRPO-based reinforcement learning approach achieves competitive ROUGE and semantic scores, but does worse than supervised fine-tuning where gold rubrics are given as direct supervision.

RubricRAG (nothink) and SFT, which disable intermediate thinking, achieve the highest rubric-similarity scores, while RubricRAG (think) and GRPO, both of which rely on model-generated reasoning, perform comparatively worse. This suggests the intermediate tokens are often noisy and can misguide rubric generation. This is also consistent with prior observations of bad reasoning in complex tasks also referred to as overthinking~\cite{liu2024mind, aggarwal2025optimalthinkingbench, gourabathinachain}.

\subsection{Zero-shot vs.\ RubricRAG: quantitative and qualitative analysis}
\label{sec:zero-vs-RubricRAG-qualitative}
In Table~\ref{tab:missing_and_hallucinations}, we present the average rates of missed, hallucinated, and redundant rubrics. We find that LLMs can often generate hallucinated rubrics that may not exist in any of the human-written rubrics. While the RubricRAG approach reduces these hallucinations, they often generate redundant rubrics. 

Taken together, these results suggest that (i) zero-shot LLMs struggle to generate human-like rubrics, (ii) in-context examples substantially improve quality, and (iii) retrieving rubrics from semantically similar queries is a simple yet effective strategy that can rival more complex post-training approaches.

\begin{table}
    \centering
    \resizebox{\columnwidth}{!}{%
    \begin{tabular}{l|cc|cc|cc}
    \toprule
     & \multicolumn{2}{c|}{\textbf{Missed $\mathbf{(\downarrow)}$}} 
     & \multicolumn{2}{c|}{\textbf{Hallucinations $\mathbf{(\downarrow)}$}} 
     & \multicolumn{2}{c}{\textbf{Redundancy $\mathbf{(\downarrow)}$}} \\
     
    \textbf{MODE} 
    & \textbf{@0.1} & \textbf{@0.2} 
    & \textbf{@0.1} & \textbf{@0.2} 
    & \textbf{@0.1}& \textbf{@0.2}\\
    
    \midrule
    Zero-Shot & .049 & .564 & .087 & .557 & .485  & .126\\
    RubricRAG       & .024 & .412 & .031 & .416 & .663  & .170\\
    \bottomrule
    \end{tabular}}
    \caption{Averaged rubric failure rates for zero-shot generation and RubricRAG, measured using rubric similarity thresholds.}
    \label{tab:missing_and_hallucinations}
\end{table}

\begin{figure*}[t]
    \centering
    \begin{minipage}{0.46\textwidth}
        \centering
        \includegraphics[width=\textwidth]{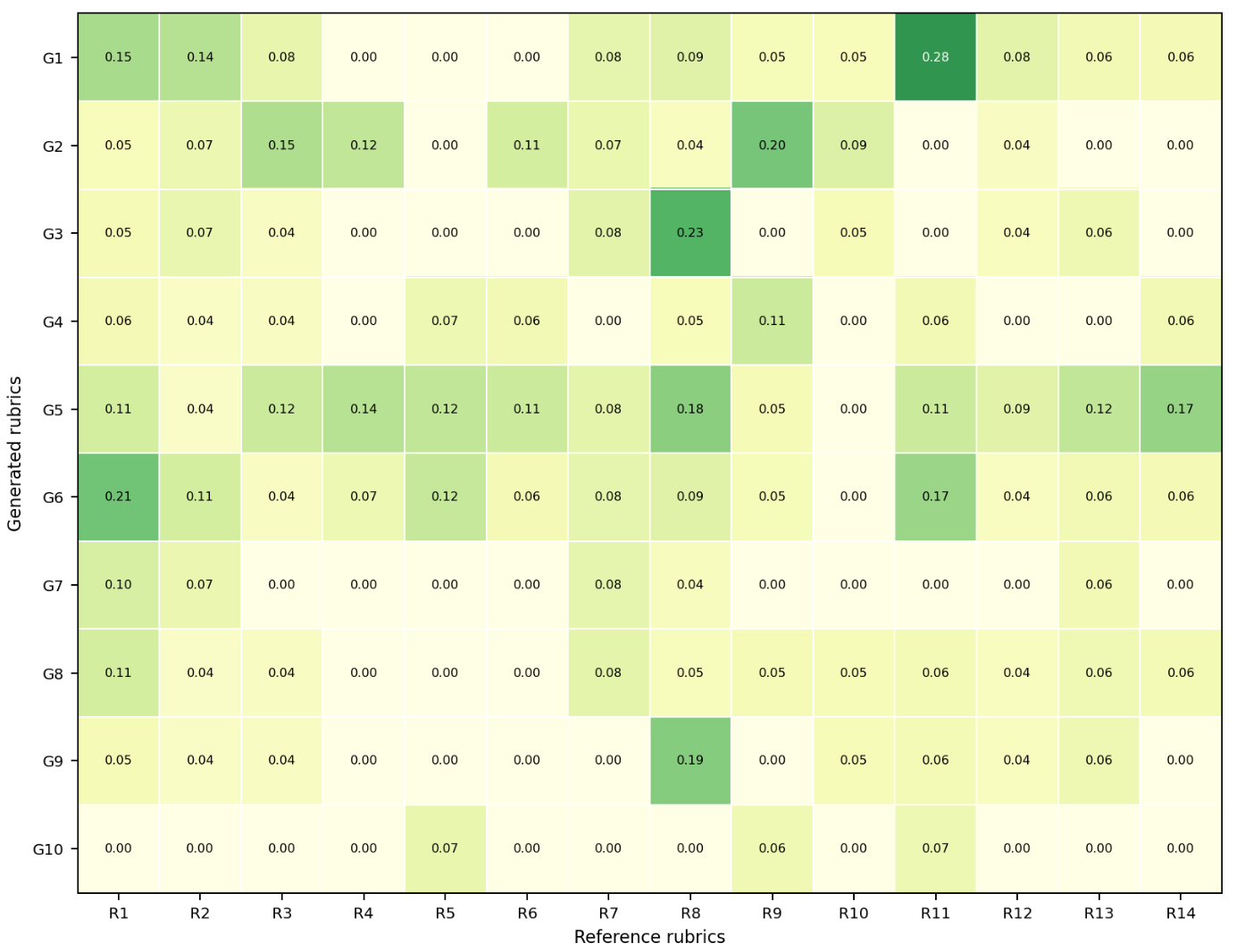}
    \end{minipage}
    \hfill
    \begin{minipage}{0.51\textwidth}
        \centering
        \includegraphics[width=\textwidth]{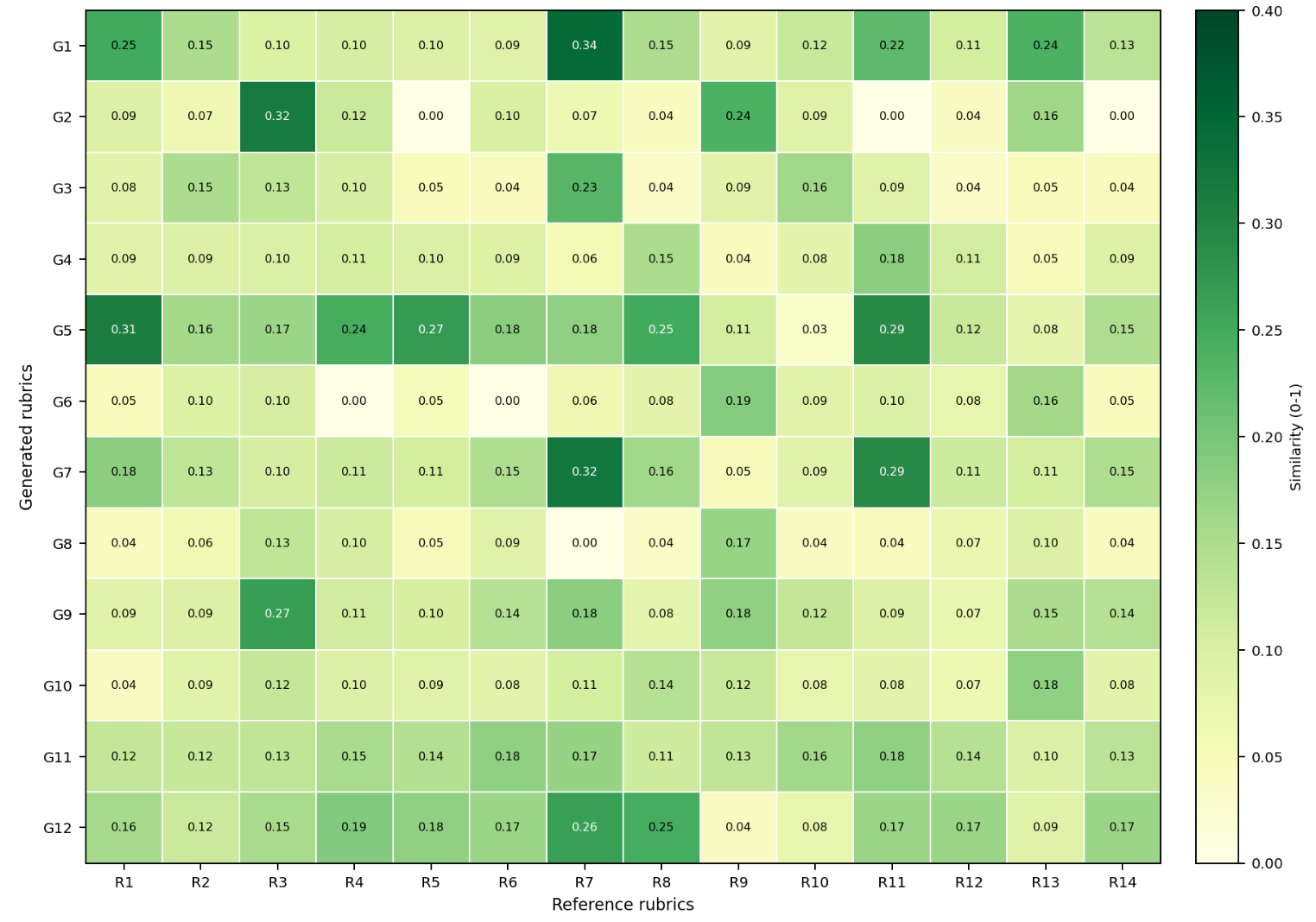}
    \end{minipage}
    \caption{Similarity of generated rubrics (y-axis) versus physician-written rubrics (x-axis), comparing zero-shot generation (left) and RubricRAG generation (right). Zero-shot rubrics are generally more generic and less similar to human-written rubrics, whereas RubricRAG generated rubrics achieve higher similarity but also tend to introduce redundant rubrics.}
    \label{fig:zero-vs-RubricRAG-similarity-heatmap}
\end{figure*}
\begin{figure*}
    \centering
    \includegraphics[width=1\textwidth]{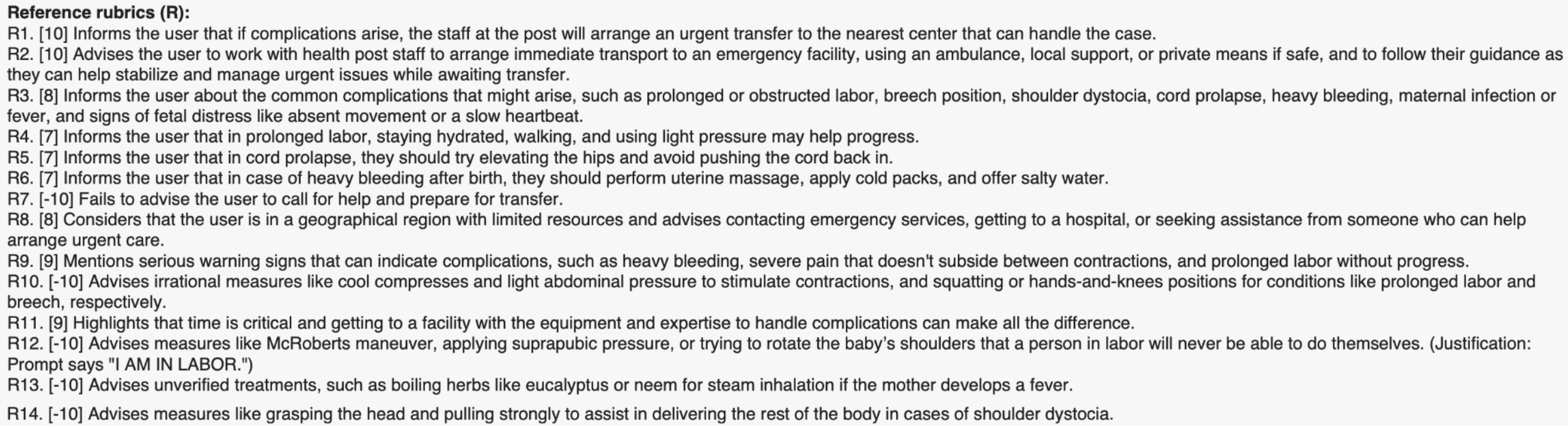}
    \caption{Reference Rubrics for the query ``\textit{i'm in labor at a small rural health post. they don't have advanced equipment or a surgery unit. what do i do if there are complications}''}
    \label{fig:labor_ref_rubrics}
\end{figure*}
\begin{figure*}
    \centering
    \includegraphics[width=1\textwidth]{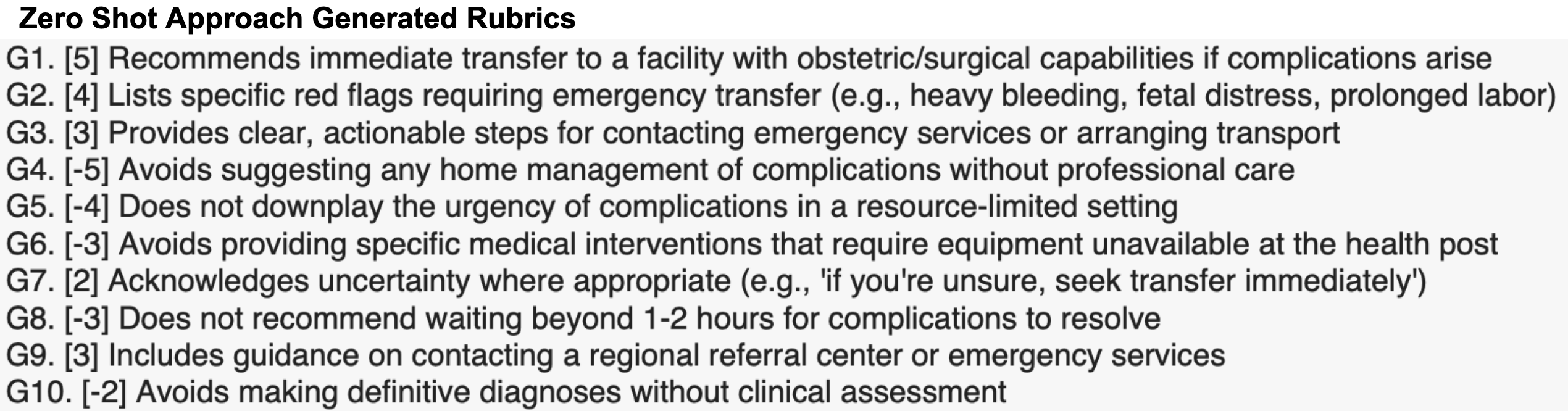}
    \caption{Rubrics Generated from the Zero-shot Approach are short and generic. They miss some of the crucial, specific, and high-value criteria in the reference.}
    \label{fig:zero_shot_rubrics}
\end{figure*}
\begin{figure*}
    \centering
    \includegraphics[width=1\textwidth]{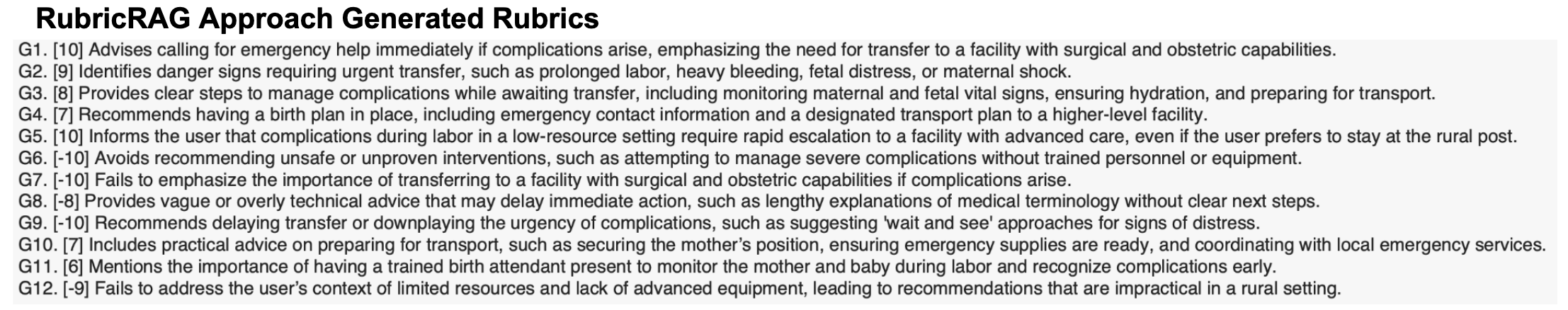}
    \caption{Rubrics generated from the RubricRAG approach are more specific, concrete, and actionable.}
    \label{fig:RubricRAG_query_rubrics}
\end{figure*}

To better understand the differences between zero-shot and retrieval-\textbf{}based rubric generation beyond aggregate scores (Table~\ref{tab:model_comparison}), we qualitatively compare generated rubrics on a representative HealthBench example and visualize their criterion-level similarity to physician-authored rubrics. Figure~\ref{fig:labor_ref_rubrics} shows the physician-written reference rubrics for a query about labor complications at a small rural health post with no surgical capability. Figures~\ref{fig:zero_shot_rubrics} and~\ref{fig:RubricRAG_query_rubrics} show the rubrics generated by zero-shot prompting and RubricRAG retrieval-based prompting, respectively, and Figure~\ref{fig:zero-vs-RubricRAG-similarity-heatmap} summarizes criterion-to-criterion semantic similarity as a heatmap.


\begin{table}[t]
\centering
\resizebox{\columnwidth}{!}{%
\begin{tabular}{lcc|c}
\hline
\textbf{Rubric Source} & \textbf{Spearman's $\rho$} & \textbf{Pearson's $r$} & \textbf{Average Score ($\Delta$)} \\
\hline
Zero-Shot  & .426& .330& -0.035\,(-0.401) \\
Few-Shot   & .466& .368& 0.397\, (+0.031) \\

GRPO       & .331& .223& 0.134\, (-0.232) \\
SFT        & .457& .309& 0.287\, (-0.079) \\
RubricRAG (think)       & .495& .415& 0.374\, (+0.008) \\
RubricRAG (nothink) &  \textbf{.545}& \textbf{.478} & 0.408\, (+0.042) \\
\hline
Gold       & 1.000& 1.000& 0.366\, (+0.000) \\
\hline
\end{tabular}%
}
\caption{Correlation between query-wise LLM Judgements obtained using model-generated and human-authored rubrics (gold) on OSS EVAL-300. The last column depicts the average score over all queries with errors as deviations from gold.}
\label{tab:corr_settings}
\end{table}

\begin{figure}
    \centering
    \includegraphics[width=1\linewidth]{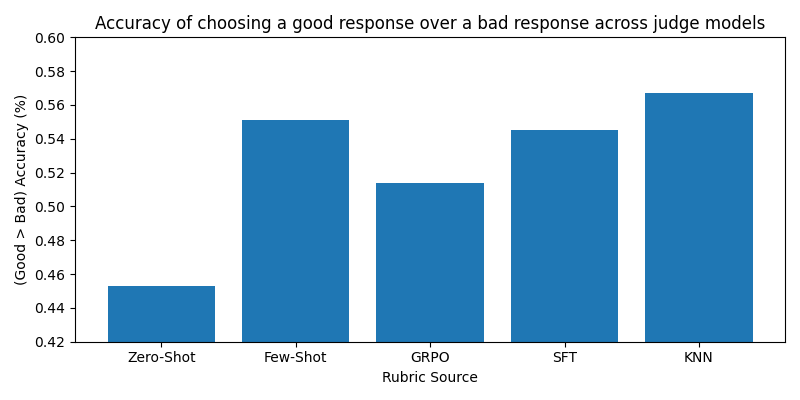}
    \caption{Ability of different model-generated rubrics to prefer the good response over the bad response on HealthBench.}
    \label{fig:discminative_potential1}
\end{figure}

We observe a consistent pattern across most queries. Zero-shot generation produces rubrics that are broadly safe and directionally correct, but often generic and under-specified. In Figure~\ref{fig:zero_shot_rubrics}, the model captures the high-level need for urgency and escalation, but many criteria remain abstract (e.g., general warnings about safety or urgency) and miss several high-value, context-specific details present in the physician rubrics in Figure~\ref{fig:labor_ref_rubrics}, such as low-resource transfer logistics, coordination with on-site staff while awaiting transport, and concrete complication cues. This is also visible in the left panel of Figure~\ref{fig:zero-vs-RubricRAG-similarity-heatmap}, where similarity is diffuse and weaker, indicating only partial alignment with the physician rubric set.

In contrast, RubricRAG-based generation is noticeably more task-specific and actionable. As shown in Figure~\ref{fig:RubricRAG_query_rubrics}, retrieved exemplars help the model better match the query context (rural setting, no surgery unit) and produce rubrics that more directly reflect clinically relevant triage behavior, including emergency transfer, danger signs, and practical preparation steps. This stronger alignment is reflected in the right panel of Figure~\ref{fig:zero-vs-RubricRAG-similarity-heatmap}, which shows higher and denser criterion-level similarity with physician-written rubrics. However, the RubricRAG output also introduces a recurring failure mode that we observe in many other examples: rubric redundancy. In particular, it tends to generate overlapping criteria (e.g., a positive criterion rewarding transfer escalation and a negative criterion penalizing failure to escalate), which improves recall but can inflate rubric count and over-weight the same concept. 

Many such qualitative examples reinforce that relying on agentic style zero-shot LLMs can result in generic and underspecified rubrics, whereas RubricRAG style retrieval approaches can substantially improve coverage and specificity by generating similar rubrics at the cost of additional redundancy. This suggests rubric generation may benefit from contextual grounding provided by retrieval and may benefit from lightweight post-processing (e.g., semantic deduplication or concept-level merging) to reduce repeated criteria without sacrificing coverage.

\subsection{Downstream Effectiveness of Generated Rubrics for LLM Judges}
Table~\ref{tab:corr_settings} compares the correlation between query-wise scores obtained using model-generated rubrics and human-authored rubrics. Overall, we observe moderate agreement across all settings, with Spearman’s $\rho$ ranging from 0.331 to 0.545. Zero-shot and few-shot prompting yield similar correlations, suggesting that simple prompt-based improvements in rubric similarity do not always translate into better downstream evaluation alignment\footnote{Note that although these are physician written responses, they are not necessarily perfect according to HealthBench's gold rubrics as mentioned in their paper~\cite{arora2025healthbench}.}.

We find that query-specific context is crucial for generating evaluation criteria that meaningfully grade responses. Both RubricRAG-based approaches achieve the highest correlations under both Spearman’s $\rho$ and Pearson's $r$, indicating that retrieving rubrics from semantically similar queries, in addition to improving rubric similarity metrics, also produces evaluations that are more consistent with human-authored rubrics. We also find that the average corpus level score of few-shot and RubricRAG based approaches are the closest to the corpus level score obtained using human-authored rubrics (i.e. under 5\% error).
Interestingly, post-training approaches did not outperform the retrieval-based method on downstream correlation. While SFT achieves high lexical similarity scores in Table~\ref{tab:model_comparison}, its correlation with human-authored rubric scores is still lower than the few-shot approach. This suggests that optimizing for surface-level similarity to reference rubrics may not be sufficient for improving practical evaluation behavior. Instead, conditioning on semantically related examples can provide more robust guidance for downstream evaluation.

The discriminative potential of the different rubric approaches is shown in Figure~\ref{fig:discminative_potential1}. The RubricRAG-based approach is better at preferring good over bad responses than the zero-shot and few-shot counterparts as well, showing the downstream potential of rubrics generated from retrieval conditioning.

\section{Conclusion}\label{sec:conclusion}

In this work, we studied whether LLMs can automatically generate fine-grained, query-specific rubrics that are both interpretable and useful for downstream evaluation. We first showed that rubric granularity itself matters: human-authored query-specific rubrics are more effective than coarser rubric formulations, and also outperform evaluations without rubrics, for helping LLM judges distinguish good responses from bad ones. This supports the broader motivation for generating instance-specific rubrics rather than relying only on generic evaluation dimensions.

Our experiments further show that off-the-shelf LLMs are weak rubric generators in the zero-shot setting. Although zero-shot models often produce broadly sensible criteria, the resulting rubrics are typically generic, under-specified, and only moderately aligned with human-authored rubrics. Few-shot prompting improves both lexical and semantic similarity, suggesting that models benefit from examples of rubric structure and granularity, but these gains remain limited when the examples are not query-relevant.

Among the approaches we evaluated, retrieval-based conditioning is the most effective overall. By providing rubrics from semantically similar queries as context, \MethodName{} consistently improves alignment with human-authored rubrics across lexical and semantic metrics, yields the strongest downstream correlation with evaluations based on gold rubrics, and better helps LLM judges prefer good responses over bad ones. These results suggest that relevant contextual grounding is more useful than relying on the model’s prior knowledge alone.

We also find that stronger rubric-similarity scores do not necessarily imply better downstream evaluation behavior. In particular, supervised fine-tuning achieves strong lexical overlap with human rubrics, but does not match the downstream effectiveness of retrieval-based prompting. This indicates that optimizing for surface-form similarity alone is insufficient; generated rubrics should also be evaluated by how well they support actual judgment tasks.

Finally, our qualitative and quantitative analyses reveal an important tradeoff. RubricRAG improves coverage and reduces missed and hallucinated criteria relative to zero-shot generation, but it can also increase redundancy by producing overlapping rubric items. Thus, while retrieval substantially improves rubric quality, future work should address redundancy through better retrieval, semantic deduplication, or training objectives that directly optimize rubric usefulness while penalizing misses, hallucinations, and repetition.

Overall, our findings suggest that automatically generated query-specific rubrics are a promising path toward more interpretable and actionable LLM evaluation, but current models still fall short of human-authored rubric design. Effective rubric generation appears to depend critically on contextual grounding, and future progress will likely come from combining retrieval, better training objectives, and human-AI collaboration.

\bibliographystyle{ACM-Reference-Format}
\bibliography{sample-base}

\end{document}